\documentclass[preprint,showpacs,showkeys,preprintnumbers,aps,epsf,amssymb]{revtex4}
\usepackage{graphicx}

\def\ben{\begin{equation}}
\def\een{\end{equation}}
\def\bey{\begin{eqnarray}}
\def\eey{\end{eqnarray}}
\def\ba{\begin{array}}
\def\ea{\end{array}}

\def\lgl{\langle}
\def\rgl{\rangle}

\def\psla{p{\raise1pt\hbox{$\!\!/$}}}
\def\dsla{\partial{\raise1pt\hbox{$\!\!\!/$}}}
\def\Dsla{D{\raise1pt\hbox{$\!\!\!/$}}}
\def\xsla{x{\raise1pt\hbox{$\!\!\!/$}}}

\def\jmu5{j_{\mu 5}^{(i)}(0)}
\def\jnu5{j_{\nu 5}^{(i)}(0)}

\def\gA#1{g_{A}^{({#1})}}

\def\qqv{\langle{\bar q}q\rangle_{0}}
\def\uuv{\langle{\bar u}u\rangle_{0}}

\def\ddv{\langle{\bar d}d\rangle_{0}}

\def\ssv{\langle{\bar s}s\rangle_{0}}

\def\qq0v{\langle0\!\mid\!{\bar q}q\!\mid\! 0\rangle}

\def\qc0f{\langle0\!\mid\!{\bar q}q\!\mid\!0\rangle_{F}}

\def\qsq0f{\langle0\!\mid\!{\bar q}\sigma_{\mu\nu}q\!\mid\!0\rangle_{F}}

\def\qgdq0f{\langle0\!\mid\!{\bar q}{\cal 
S}\gamma_{\mu}D_{\nu}q\!\mid\!0\rangle_{F}}

\def\qddq0f{\langle0\!\mid\!{\bar q}{\cal
S}D_{\mu}D_{\nu}q\!\mid\!0\rangle_{F}}
\def\m#1{m_{#1}}
\def\N#1{\langle{\bar {#1}}{#1}\rangle_{N}}
\def\ggN{\left\langle\frac{\alpha_s}{\pi}G^2\right\rangle_{N}}
\def\gdN#1{i\left\langle{\bar {#1}}{\cal 
S}\gamma_{\mu}D_{\nu}{#1}\right\rangle_{N}}
\def\gdddN#1{i\left\langle{\bar {#1}}{\cal 
S}\gamma_{\mu}D_{\nu}D_{\lambda}D_{\sigma}{#1}\right\rangle_{N}}
\def\sgmtrm{\Sigma_{\pi N}}
\def\mmnt#1#2{A_{#1}^{#2}(\mu^2)}
\def\mmntQSR#1#2{A_{#1}^{#2}(1\,{\rm GeV}^2)}
\def\3mmtm{|{\bf q}|^2}

\def\eq#1{Eq.(\ref{#1})}
\def\eqs#1#2{Eqs.(\ref{#1}) and (\ref{#2})}

\def\Ref#1{[\ref{#1}]}
\def\Refs#1#2{[\ref{#1},\ref{#2}]}

\def\nnbr{\nonumber}
\def\p0{p_0}

\def\gam3{\mbox{\boldmath{$\gamma$}}}
\def\e0{E_{0}(s_{0},s)}
\def\e1{E_{1}(s_{0},s)}
\def\e2{E_{2}(s_{0},s)}

\def\aplt{\kern0.3333em \raise 0.2ex \hbox{$<$}%
\kern-0.8em \lower0.8ex \hbox{$\sim$}%
\kern0.3333em}

\def\Gam#1#2{\Gamma^{#1}_{#2}}
\def\msbar{\overline{\rm MS}}

\begin{document}
\preprint{KEK-TH-967}

\title{Flavour-singlet $g_A$ and the QCD
 sum rule incorporating instanton effects}
\author{Tetsuo NISHIKAWA}
\email{nishi@post.kek.jp}
\affiliation{Institute of Particle and Nuclear Studies, High Energy
Accelerator Research Organization, 1-1, Ooho, Tsukuba, Ibaraki, 305-0801, Japan}

\date{\today}

\baselineskip 18pt
\begin{abstract}
We derive a QCD sum rule for the flavour-singlet axial 
coupling constant $g_A^{(0)}$ 
from a two point correlation function of
flavour-singlet axial vector currents in a one-nucleon state.
In evaluating the correlation function by an operator product expansion
we take into account the terms up to dimension 6.
This correlation function receives an additional two-loop diagram
which comes from an (anti-)instanton.
If we do not include it, $g_A^{(0)}$ is estimated to be 0.8.
However, the additional diagram due to instantons contributes negatively
and reduces $g_A^{(0)}$ towards the experimental value.
\end{abstract}

\maketitle

\newpage
\indent

Axial coupling constants are defined by
nucleon matrix elements of axial currents at zero momentum transfer.
Since an axial current, ${\bar q}(x)\gamma_{\mu}\gamma_5 q(x)$,
is a spin operator,
the flavour-singlet axial coupling constant $\gA{0}$
represents the fraction of the nucleon spin carried by quarks. 
In the naive parton model, $\gA{0}$ is expected to be close to
$1$.
However, an unexpected 
small value of $\gA{0}$ was found from the EMC experiment,
which implies the quarks contribute only a small fraction to the
proton's spin.
This has led to the so-called \lq\lq spin crisis''
and raised a number of questions of understanding the dynamics of
the proton spin [\ref{Anselmino}].
A number of subsequent experiments have been performed.
The results are in the range $\gA{0}=(0.28-0.41)$, see [\ref{Ji}] for a recent review.

The investigations of $\gA{0}$ by QCD sum rules 
have been done so far by the authors in 
Refs.[\ref{Ioffe},\ref{Belitsky}].
Ioffe and Oganesian \Ref{Ioffe} derived a QCD sum rule for $\gA{0}$
by considering a two-point correlation function of 
nucleon interpolating fields in an external flavour-singlet axial-vector field.
In their sum rule, $\gA{0}$ is expressed by expectation values of
QCD composite operators induced by the external field.
An important contribution comes from dimension 3 term in the
operator product expansion, which is related to the derivative
of QCD topological susceptibility $\chi'(0)$.
They found the lower limit of $\gA{0}$ and $\chi'(0)$ from the
self consistency of the sum rule: $\gA{0}\geq 0.05$,
$\chi'(0)\geq 1.6\times 10^{-3} {\rm GeV}^2$.

The authors in Ref.\Ref{Belitsky}
considered a three-point function of nucleon 
interpolating fields and the divergence of a flavour-singlet axial-vector current.
They took into account chiral anomaly by using the anomaly relation
from the very beginning.
The form factor, $g_{A}^{(0)}(q^2)$, is related to 
the vacuum condensates of the 
quark-gluon composite operators through a double dispersion relation. 
To know $g_{A}^{(0)}(q^2)$ at $q^2=0$ one must evaluate the correlation
function at zero momentum.
Although the method to evaluate it is known \Ref{Balitsky}, 
it involves large uncertainty.

Recently, we have proposed a new method to construct QCD sum rules for
axial coupling constants from two-point correlation functions 
of axial-vector currents in a one-nucleon state \Ref{TN}.
With the method, the axial coupling constants are expressed in
terms of the $\pi$-$N$ and $K$-$N$ sigma-terms and
the moments of parton distributions.
We have seen good agreement with experiment for the non-singlet constants, $g_{A}^{(3)}$
and $g_{A}^{(8)}$.

In this paper, we extend the previous work to the case for  $\gA{0}$. 
For the calculation of $\gA{0}$ 
we need to fully take into account the chiral anomaly.
Since the origin of the chiral anomaly is considered to be instantons,
one might suspect the anomalous suppression of $\gA{0}$
is somehow related to instantons.
We therefore evaluate their effects on $\gA{0}$.


Following Ref.\Ref{TN}, we consider a correlation function of
flavour-singlet axial-vector currents in a one-nucleon state:
\ben
\Pi^{(0)}_{\mu\nu}(q;P)
=i\int d^4 x {\rm e}^{iqx}\lgl T[j_{\mu 5}^{(0)}(x),j_{\nu 5}^{(0)}(0)]\rgl_N,
                                  \label{corfun}
\een
where $q^{\mu}\equiv(\omega,{\bf q})$ and 
the nucleon matrix element is defined by
$\lgl\ldots \rgl_N \equiv (1/2)\sum_S
\left[
\lgl N(PS)|\ldots |N(PS)\rgl-\lgl\ldots\rgl_0 \lgl N(PS)|N(PS)\rgl
\right]$,
where $P^{\mu}\equiv(E,{\bf P})$ is the nucleon momentum ($P^2=M^2$,
$M$ is the mass), $S$ the nucleon spin, $\lgl\ldots\rgl_0
\equiv \lgl 0|\ldots |0\rgl$, 
and the one-nucleon state is normalized as $\lgl N(PS)|N(P'S')\rgl=(2\pi)^3 
\delta^{3}({\bf P}-{\bf P'})\delta_{SS'}$. 
The flavour-singlet axial-vector current is defined as
\ben
j_{\mu 5}^{(0)}(x)=
{\bar u}(x)\gamma_{\mu}\gamma_{5}u(x)+{\bar 
d}(x)\gamma_{\mu}\gamma_{5}d(x)
+{\bar s}(x)\gamma_{\mu}\gamma_{5}s(x),          \label{jmu50}
\een
where $u$, $d$ and $s$ are the up, down and strange quark fields,
respectively.

We write a Lehmann representation of \eq{corfun}:
\ben
\Pi^{(0)}_{\mu\nu}(\omega,{\bf q};P)=\int^{\infty}_{-\infty}
d\omega'\frac{\rho^{(0)}_{\mu\nu}(\omega',{\bf 
q};P)}{\omega-\omega'},\label{Lehmann}
\een
where $\rho_{\mu\nu}$ is the spectral function.
We derive a Borel sum rule from \eq{Lehmann} for the even function part of
\eq{corfun}
in $\omega$, $\Pi^{(0)}_{\mu\nu}(\omega^2,{\bf q};P)_{\rm even}$, 
as \Ref{TN}
\ben
\widehat{B}[\Pi^{(0)}_{\mu\nu}(\omega^2,{\bf q};P)_{\rm even}]
=-\int^{\infty}_{-\infty}
d\omega'\omega' \exp\left(-\omega'^2/s\right)\rho^{(0)}_{\mu\nu}(\omega',{\bf q};P),
\label{BSR-even}
\een
where $s$ denotes the square of the Borel mass 
and $\widehat{B}$ the Borel transformation with respect to $\omega^2$.
In \eq{BSR-even} the left hand side is evaluated theoretically,
which give rise to a Borel transformed 
QCD sum rule.

Let us now consider the physical content of the spectral function
with the insertion of intermediate states between the currents.
Here the lowest one is a one-nucleon state.
The continuum state consists of $\eta'$-nucleon states, excited
nucleon states and so on.
There is an energy gap between the nucleon pole and the continuum
threshold.
The contribution of the one-nucleon state to the spectral function is 
expressed in terms of axial coupling constants,
because the nucleon matrix element of $j_{\mu 5}^{(0)}$
is written as $\lgl N(PS)|j_{\mu 5}^{(0)}(0)|N(P'S')\rgl=
{\bar u}(PS)\left[g_{A}^{(0)}(q^2)\gamma_{\mu}\gamma_5
+h_{A}^{(0)}(q^2)
q_{\mu}\gamma_5 \right]u(P'S')$,
where $u(PS)$ is a Dirac spinor and
$q=P'-P$ \Ref{TN}.
The contribution of the continuum state becomes small in the Borel sum rule,
since it is exponentially suppressed compared to that of the 
one-nucleon state because of the energy gap.
Therefore, it is allowed to use a rough model of the continuum:
The form of the continuum is approximated 
by the step function with the coefficient 
being the imaginary part of the asymptotic form 
of the correlation function in the 
OPE \Ref{SVZ1}.
In the present case, however, the continuum contribution to the 
spectral function is absent within the approximation,
because the perturbative part is subtracted from 
the definition of \eq{corfun}.
This means that the continuum contribution may be very small
at least in the high energy region.
We therefore neglect the continuum contribution 
in this work.

Hereafter we consider 
the correlation function in the rest 
frame of the initial and final nucleon states and contract 
the Lorentz indices of the currents.
Expanding the right hand side of \eq{BSR-even} in powers of $\3mmtm$,
we find the coefficient of $|{\bf q}|^2$ is proportional to
$|g^{(0)}_{A}(0)|^2$ \Ref{TN}.
$h_{A}^{(0)}(0)$ contributes to higher order terms 
since $h_{A}^{(0)}(q^2)$ has no singularity at $q^2=0$.
From the first derivative of \eq{BSR-even} with respect to $\3mmtm$
we obtain the desired QCD sum rule at $|{\bf q}|^2=0$ : 
\ben
\left.\frac{\partial\hat{B}\left[\Pi^{(0)}(\omega^2,{\bf q})_{\rm even}\right]
}{\partial|{\bf q}|^2} \right|_{|{\bf q}|^2=0}
=-\frac{3}{M}|g^{(0)}_{A}|^2
, \label{BSR0}
\een
where $\Pi(\omega,{\bf q})={\Pi_\mu}^\mu(\omega,{\bf q};M,{\bf 0})$.


Let us now turn to the evaluation of $\Pi^{(0)}(q)$.
$\Pi^{(0)}(q)$ consists of the following two terms:
\ben
\Pi^{(0)}(q)=\sum_{q=u,d,s}C(q)_{q}+\sum_{q,q'=u,d,s}D(q)_{qq'}, 
\label{contract-Pi}
\een
where $C(q)_{q}$ is a connected or \lq\lq
one-loop'' term which is given by
\ben
C(q)_{q}=
-i\int d^4 (x-y){\rm e}^{iq(x-y)}\left\langle{\rm Tr}
\left\{T\left[q(y){\bar q}(x)\right]
\gamma_{\mu}\gamma_5 T\left[q(x){\bar q}(y)\right]
\gamma^{\mu}\gamma_5\right\}
\right\rangle_N
\label{1loop}
\een
and $D(q)_{qq'}$ a disconnected or \lq\lq two-loop'' term: 
\ben
D(q)_{qq'}=i\int d^4 (x-y){\rm e}^{iq(x-y)}
\left\langle{\rm Tr}\left\{\gamma_{\mu}\gamma_5
 T\left[q(x){\bar q}(x)\right]\right\}{\rm Tr}\left\{\gamma^{\mu}\gamma_5
T\left[q'(y){\bar q'}(y)\right]\right\}
\right\rangle_N. \label{2loop}
\een
Note here that {\it only the correlator of flavour singlet
currents receives the contributions of the \lq\lq two-loop'' terms}.
Indeed, for the correlation function of iso-vector currents,
$j_{\mu5}^{(3)}=(1/2)({\bar u}\gamma_{\mu}\gamma_5 u-{\bar
d}\gamma_{\mu}\gamma_5 d)$,
\lq\lq two-loop'' terms cancel with each other if we neglect the
differences of quark masses: 
\begin{eqnarray}
\Pi^{(3)}(q)&\equiv&i\int d^4 x {\rm e}^{iqx}\lgl T[j_{\mu 5}^{(3)}(x),j_{\nu 5}^{(3)}(0)]\rgl_N
\nnbr\\
&=&\left(1/2\right)^2[C_{u}(q)+C_{d}(q)+D(q)_{uu}-2D(q)_{ud}+D(q)_{dd}]\nnbr\\
&=&\left(1/2\right)^2[C_{u}(q)+C_{d}(q)].
\eey
Similarly, for the 8-th component of flavour-octet currents,
$j_{\mu5}^{(8)}=(1/2\sqrt{3})({\bar
u}\gamma_{\mu}\gamma_5 u
+{\bar d}\gamma_{\mu}\gamma_5 d-2{\bar s}\gamma_{\mu}\gamma_5 s)$,
\bey
\Pi^{(8)}(q)&\equiv&i\int d^4 x {\rm e}^{iqx}\lgl T[j_{\mu 5}^{(8)}(x),j_{\nu 5}^{(8)}(0)]\rgl_N \nnbr\\
&=&\left(1/2\sqrt{3}\right)^2[C_{u}(q)+C_{d}(q)+4C_{s}(q)\nnbr\\
&&+D(q)_{uu}+D(q)_{dd}+4D(q)_{ss}+2D(q)_{ud}-4D(q)_{us}-4D(q)_{ds}]\nnbr\\
&=&\left(1/2\sqrt{3}\right)^2[C_{u}(q)+C_{d}(q)+4C_{s}(q)].
\end{eqnarray}

We evaluate \eqs{1loop}{2loop} by a standard operator product expansion (OPE).
Let us first consider the \lq\lq one-loop'' terms. 
In the OPE, operators of the leading terms are of dimension 4.
We take into account the terms up to dimension 6.
The result is in the following:
\bey
C(q)_{q}&=&
\frac{10}{q^2}\m{q}\N{q}
-\frac{1}{2q^2}\ggN-\frac{8q^{\mu}q^{\nu}}{q^4}\gdN{q} \nnbr \\
& &-\frac{22\pi\alpha_s}{3q^4}
      \left\langle{\bar q}\gamma^{\mu}\lambda^{a}q
                  \left({\bar u}\gamma_{\mu}\lambda^{a}u
                       +{\bar d}\gamma_{\mu}\lambda^{a}d
                       +{\bar s}\gamma_{\mu}\lambda^{a}s\right)
      \right\rangle_{N} \nnbr \\
& &+\frac{10\pi\alpha_s q^{\mu}q^{\nu}}{q^6}
      \left\langle{\cal S}\left({\bar q}\gamma_{\mu}\lambda^{a}q\right)
                  \left({\bar u}\gamma_{\nu}\lambda^{a}u
                       +{\bar d}\gamma_{\nu}\lambda^{a}d
                       +{\bar s}\gamma_{\nu}\lambda^{a}s\right)
      \right\rangle_{N} \nnbr \\
& &+\frac{32 q^{\mu}q^{\nu}q^{\lambda}q^{\sigma}}{q^8}\gdddN{q}  \label{OPE0}
\eey
where $D_\mu$'s are covariant derivatives,
$G^2\equiv G^a_{\mu\nu}G^{a\mu\nu}$, and ${\cal S}$ denotes a symbol which
makes the operators symmetric and traceless with respect to the Lorentz
indices.

We now discuss about the nucleon matrix elements in \eq{OPE0}.
It is known well that $m_{q}\N{q}$ is related to the $\pi$-$N$ or
$K$-$N$ sigma-term as
$(\m{u}+\m{d})(\N{u}+\N{d})=2\sgmtrm$ and 
$(\m{s}+\m{u})(\langle\bar ss\rangle_N+\langle\bar uu\rangle_N)
=2\Sigma_{KN}$.
$\lgl(\alpha_s /\pi)G^2\rgl_{N}$ 
is expressed by the nucleon mass and $m_{q}\N{q}$ through the
QCD trace anomaly:$
\lgl(\alpha_s /\pi)G^2\rgl_{N}=-(8/9)\Big(M-\sum_{q=u,d,s} m_{q}\N{q}\Big)$.
The matrix elements which contain covariant derivatives are related to 
the parton distributions as
$\langle{\cal S}{\bar q}\gamma_{\mu_{1}}D_{\mu_{2}}\cdots
D_{\mu_{n}}q(\mu^2)\rangle_{N}
=(-i)^{n-1}A_{n}^{q}(\mu^2)T_{\mu_{1}\mu_{2}\ldots \mu_{n}}$,
where $A_{n}(\mu^2)$ is the $n$-th moment of the parton distributions 
at scale $\mu^2$, 
and $T_{\mu_{1}\mu_{2}\ldots \mu_{n}}={\cal S}\left[P_{\mu_{1}}P_{\mu_{2}}\cdots 
P_{\mu_{n}}\right]$.
For the matrix elements of four quark operators,
we apply the factorization hypothesis \Ref{SVZ1}:
the matrix elements are factorized by assuming that the
contribution from one nucleon state dominates in the intermediate
states:
$\lgl{\cal O}_1 {\cal O}_2 \rgl_N\approx
\lgl{\cal O}_1 \rgl_N \lgl{\cal O}_2 \rgl_0 +\lgl{\cal O}_1 \rgl_0 \lgl{\cal 
O}_2 \rgl_N $. 
We apply this hypothesis to the following type of the
nucleon matrix elements,
which appear in \eq{OPE0}:
$\lgl{\bar q}_{f}\gamma_{\mu}\lambda^a q_{f}
{\bar q}_{f'}\gamma_{\nu}\lambda^a q_{f'}\rgl_{N}
=-(8/9)g_{\mu\nu}\lgl{\bar q}_{f}q_{f}\rgl_{0}
\lgl{\bar q}_{f}q_{f}\rgl_{N}\delta_{f,f'}$,
where $f$ and $f'$ are flavor indices.

Let us next consider the \lq\lq two-loop term'', \eq{2loop}.
There is no contribution from this term as long as
we do not account for higer dimensional terms (larger than dimension 6)
in the OPE and perturbative corrections.
However, \lq\lq two-loop term" receives a contribution arising 
from instantons.
We evaluate it under the \lq\lq dilute instanton gas
approximation'' (DIGA) \Ref{Shuryak}. 
Namely we assume there exists only one
instanton or anti-instanton in vacuum.
We have two reasons why this approximation is expected to be valid.
The first is that since we use the framework of QCD sum rules we
are interested only in the short distance behavior of the
correlation function.
The second is that instantons in vacuum is sufficiently dilute \Ref{Shuryak}.
Indeed, it is known that the density of
instantons is about $1\,{\rm fm}^{-4}$ 
and the critical size of an (anti-)instanton $\rho_c$
is $\rho_c\simeq 0.3\,{\rm fm}$.
The value of $\rho_c$ is significantly smaller than the
typical separation between instantons.

In order to evaluate \eq{2loop},
we first consider the correlation function in nuclear matter with
its baryon number density being $\rho_B$.
Then \eq{2loop} is obtained as the first derivative of the
correlator in nuclear matter with respect to $\rho_B$,
because, in general, expanding a vacuum expectation value 
of an operator ${\cal O}$ at finite
baryon number density in powers of $\rho_B$ the coefficient of 
the linear term 
is nothing but the nucleon matrix element:
$\langle{\cal O}\rangle_{\rho_B}=\langle{\cal O}\rangle_{0}+\rho_B 
\langle{\cal O}\rangle_{N}+\cdots$.
Thus \eq{2loop} can be written as
\ben
D(Q)_{qq'}=\left[\frac{\partial}{\partial\rho_B}\int d^4 (x-y) {\rm e}^{iQ(x-y)}
\langle{\rm Tr}\left\{\gamma_{\mu}\gamma_{5}T\left[q(x){\bar
q}(x)\right]\right\}
{\rm Tr}\left\{\gamma^{\mu}\gamma_{5}
T\left[q'(y){\bar q'}(y)\right]\right\}
\rangle_{\rho_B}
\right]_{\rho_B=0}, \label{Piqq'E}
\een
where $Q$ is an Euclidean momentum defined by $Q^2=-q^2$ and
$\langle\cdots\rangle$ means that averaging is performed over all gauge
configurations with the weight function ${\rm exp}(-S)$,
where $S$ is the Euclidean action.
Under DIGA \eq{Piqq'E} becomes [\ref{Andrei}] 
\bey
D(Q)_{qq'}&=&\int d\rho
\left[\frac{\partial}{\partial\rho_B}n(\rho,\rho_B)\right]_{\rho_B=0}
\nnbr\\
&&\times
\int d^4 x {\rm e}^{iQx}
{\rm Tr}\left[\gamma_{\mu}\gamma_{5}S_{q}(x,x)\right]
\int d^4 y  {\rm e}^{-iQy}
{\rm Tr}\left[\gamma_{\nu}\gamma_{5}S_{q'}(y,y)\right], \label{Pi-factrzd}
\eey
where 
$S_q(x,y)$ is a quark (q) propagator in the field of an
(anti-)instanton.
We performed integrations over the center of an (anti-)instanton $z$ and
instanton size $\rho$ with a weight $n(\rho,\rho_B)$.
$n(\rho,\rho_B)$ is the tunneling rate at finite baryon
number density.

$S_q(x,y)$ is expressed as follows:
\begin{equation}
S_q(x,y)=\sum_{\lambda}\frac{\psi_\lambda(x)\psi_\lambda^{\dagger}(y)}{\lambda+im_{q}},
\end{equation}
where $\psi_\lambda(x)$ is an eigen function of Dirac operator
with the eigen value $\lambda$:
$\Dsla \psi_\lambda(x)=\lambda \psi_\lambda(x)$.
Then the dominant contribution comes from the zero-mode, $\psi_0 (x)$.
In \eq{Pi-factrzd}, however, quarks do not propagate 
in zero-mode states.
The reason is very simple.
A zero-mode changes its chirality in passing 
through an instanton (see Fig.\ref{fig1} (a)).
On the other hand, quarks created by an axial current 
have same chirality (Fig.\ref{fig1} (b)).
So zero-modes are not allowed in \eq{Pi-factrzd} and
only non-zero-modes contribute.
Non-zero-mode propagator $S^{NZM}_q(x,y)$, in which all
non-zero-modes are summed up,
\begin{equation}
S^{NZM}_q(x,y)\equiv \sum_{\lambda\neq 0} \frac{\psi_\lambda(x)\psi_\lambda^{\dagger}(y)}{\lambda+im_{q}},
\end{equation}
satisfies the equation
\ben
(\Dsla+im_q) S^{NZM}_q(x,y)=\delta^{(4)}(x-y)-\psi_0 (x)\psi_0^{\dagger}(y).
\een
Subtracting the zero-mode contribution in the right hand side,
the remaining $S^{NZM}_q(x,y)$ is ensured to be orthogonal
to the zero-mode [\ref{Brown}].
The solution of this equation, in general, has the following form 
[\ref{Brown}]:
\ben
S^{NZM}_q(x,y)=(\stackrel{\rightarrow}{\Dsla_{x}}-im_q)\Delta(x,y)\frac{1+\gamma_{5}}{2}
+\Delta(x,y)\stackrel{\leftarrow}{\Dsla_{y}}\frac{1-\gamma_{5}}{2},
\label{NZMprop}
\een
where $\stackrel{\rightarrow}{D_{\mu}}$ and
$\stackrel{\leftarrow}{D_{\mu}}$
are covariant derivatives: $
\stackrel{\rightarrow}{{D}_{\mu}}=\stackrel{\rightarrow}{\partial_{\mu}}
-i\frac{\tau^{a}}{2}A_{\mu}^{a}$ and $
\stackrel{\leftarrow}{{D}_{\mu}}=-\stackrel{\leftarrow}{\partial_{\mu}}
-i\frac{\tau^{a}}{2}A_{\mu}^{a}$,
with $\tau^a$'s are Pauli matrices and $A_{\mu}^{a}$ being an (anti-)instanton solution
:
\ben
A_{\mu}^{a}(x)=2\frac{x_{\nu}}{x^2}
\frac{{\bar \eta}_{a\mu\nu}\rho^2}{x^2+\rho^2}. \label{instsol}
\een
\eq{instsol} is the solution in the singular gauge.
Hereafter we work in this gauge.
An anti-instanton solution is obtained by replacing 
${\bar \eta}_{a\mu\nu}$ to $\eta_{a\mu\nu}$.
${\eta}_{a\mu\nu}$ and ${\bar\eta}_{a\mu\nu}$
are the 'tHooft symbols [\ref{Shuryak}].
$\Delta(x,y)$ is the propagator of a scalar field
which satisfies the equation
\begin{equation}
(-D^2+m_q^2)\Delta(x,y)=\delta^{(4)}(x-y).
\end{equation}
It is known that this equation is solved by
\begin{equation}
\Delta(x,y)=\frac{1}{4\pi^2(x-y)^2}\left(1+\frac{\rho^2}{x^2}\right)^{-1/2}\left(1+\frac{\rho^2}{y^2}\right)^{-1/2}
\left(1+\rho^2\frac{\tau^{-}\cdot x \tau^{+}\cdot y}{x^2
y^2}\right),
\label{scalarprop}
\end{equation}
where 
$\tau^{\pm}_{\mu}\equiv(\stackrel{\rightarrow}{\tau},\mp i)$
[\ref{Brown},\ref{Din}].
Here we have neglected ${\cal O}(m_q^2)$ terms. 
The propagator in the field of an anti-instanton is obtained
by interchanging $\tau^+$ and $\tau^-$.

The calculation of 
${\rm Tr}\left[\gamma_{\mu}\gamma_{5}S^{NZM}_q(x,x)\right]$ in \eq{Pi-factrzd}
needs some care, since $S^{NZM}_q(x,y)$ is not well defined at
$x\rightarrow y$.
This is because $S^{NZM}_q(x,y)$ at $x\neq y$ is not gauge invariant.
The trace should be defined as limit of an gauge invariant expression
in which a path ordered product is inserted:
\ben
{\rm Tr}\left[\gamma_{\mu}\gamma_{5}S^{NZM}_q(x,x)\right]
=\lim_{\epsilon\rightarrow 0}
{\rm Tr}\left\{\gamma_{\mu}\gamma_{5}S^{NZM}_q(x-\epsilon/2,x+\epsilon/2)
{\rm exp}\left[i\int_{x-\epsilon/2}^{x+\epsilon/2}
\frac{\tau^a}{2}A_{\nu}^a(z)dz_{\nu}\right]\right\}, \label{Tr-gaugeinv}
\een
where the right hand side must be averaged over all the direction
of the four vector $\epsilon_{\mu}$.
As a result we obtain
\begin{equation}
  {\rm Tr}\left[\gamma_{\mu}\gamma_5 S^{NZM}_q(x,x)\right]
=\frac{1}{4\pi^2}\frac{\partial}{\partial x_{\mu}}
\frac{2\rho^2}{(\rho^2+x^2)^2}.
\label{Tr-result}
\end{equation}
The Fourier transform of this equation is given by
\ben
\int d^4 x {\rm e}^{iQx}{\rm Tr}\left[\gamma_{\mu}\gamma_{5}S(x,x)\right]
=-i\rho^2 Q_{\mu}K_{0}(Q\rho).
\label{FT-tr}
\een
Here $K_{0}(z)$ is an 0-th modified Bessel function.
Then \eq{Pi-factrzd} reads
\ben
D_{qq'}(Q)=\int d\rho \left[\frac{\partial}{\partial\rho_B}n(\rho,\rho_B)
\right]_{\rho_B=0} Q^2\rho^4
K_{0}(Q\rho)^2.
\label{Pi-inst}
\een

An important quantity in \eq{Pi-inst} is the tunneling rate, $n(\rho,\rho_B)$.
In normal vacuum, where vacuum condensation does not exist,
the tunneling rate at one-loop was first given in Ref.\Ref{tHooft}.
After the pioneering work, it is now available in two-loop renormalization group invariant form \Ref{Schrempp}, 
\begin{eqnarray}
n(\rho)&=&n_{0}(\rho)\prod_{q=u,d,s}(m_q\rho)
(\rho\mu_0)^{n_f \gamma_0 \frac{\alpha_{\overline{\rm MS}}(\mu_0)}{4\pi}}
\label{tunrate-nv}
\end{eqnarray}
where $n_{0}(\rho)$ is that for quarkless theory:
\bey
n_{0}(\rho)&=&\frac{d_{\msbar}}{\rho^5}
\left(\frac{2\pi}{\alpha_{\msbar}(\mu_0)}\right)^{2N_c}
{\rm exp}\left(-\frac{2\pi}{\alpha_{\msbar}(\mu_0)}\right)
(\rho\mu_0)^{\beta_0 +(\beta_1 -4N_c \beta_0)\frac{\alpha_{\msbar}(\mu_0)}{4\pi}},\\
d_{\msbar}&=&\frac{2{\rm e}^{5/6}}{\pi^2 (N_c -1)!(N_c -2)!}
{\rm exp}(-1.511374 N_c +0.291746 n_f ),\\
\beta_0 &=&\frac{11}{3}N_c -\frac{2}{3}n_f,\\
\beta_1 &=&\frac{34}{3}N_c^2 -\left(\frac{13}{3}N_c -\frac{1}{N_c}\right) n_f,\\
\gamma_0 &=&3\frac{N_c^2 -1}{N_c}.
\label{tunrate-pg}
\eey
In Eqs.(\ref{tunrate-nv})$\sim$(\ref{tunrate-pg}), $\mu_0$ is some arbitrary
normalization point, $N_c =3$ and $n_f  =3$
are the numbers of color and flavor.

In physical vacuum, where vacuum condensation exists, 
according to Shifman et al. [\ref{Shifman}], 
the current quark mass in \eq{tunrate-nv}
is substituted by a dynamical one.
Shifman et al. considered the tunneling rate
for small size instantons.
Then it can be expanded in powers of $\rho$.
The coefficient of each term of the expansion 
should be a quantity characterizing 
the vacuum structure. 
They found that the the tunneling rate is written as a
vacuum expectation value of an \lq\lq effective Lagrangian", 
which has form analogous to a standard operator product expansion.
The Lagrangian is as follows:
\begin{eqnarray}
 \Delta{\cal L}&=&n_0(\rho)
 (\rho\mu_0)^{n_f \gamma_0 \frac{\alpha_{\msbar}(\mu_0)}{4\pi}}
\left\{\prod_{q=u,d,s}
\left(m_q\rho-\frac{4}{3}\pi^2\rho^3 {\bar q}\Gam{}{}q\right)
\right.\cr
&&+\frac{3}{32}\left(\frac{4}{3}\pi^2\rho^3 \right)^2
\left[\left({\bar u}\Gam{a}{}u {\bar d}\Gam{a}{}d
      -\frac{4}{3}{\bar u}\Gam{a}{\mu\nu}u 
                  {\bar d}\Gam{a}{\mu\nu}d
      \right)
      \left(m_s\rho-\frac{4}{3}\pi^2\rho^3 {\bar s}\Gam{}{}s\right)
\right.\cr
&&\left.
+\frac{9}{40}\cdot\frac{4}{3}\pi^2\rho^3 d_{abc}
 {\bar u}\Gam{a}{\mu\nu}u {\bar d}\Gam{b}{\mu\nu}d {\bar s}\Gam{c}{}s
+({\rm 2\,\,permutations})
\right]\cr
&&
\left.
+\frac{9}{320}\left(\frac{4}{3}\pi^2\rho^3\right)^3 d_{abc}
 {\bar u}\Gam{a}{}u {\bar d}\Gam{b}{}d {\bar s}\Gam{c}{}s
+\frac{9}{256}i\left(\frac{4}{3}\pi^2\rho^3\right)^3
f_{abc}{\bar u}\Gam{a}{\mu\nu}u {\bar d}\Gam{b}{\nu\gamma}d {\bar s}\Gam{c}{\gamma\mu}s
\right\}, \label{efflag}\cr
&&\end{eqnarray}
where $\Gam{}{}=(1-\gamma_5)/2,\quad
 \Gam{a}{}=(1-\gamma_5)/2\cdot(\lambda^a/2),\quad
 \Gam{a}{\mu\nu}=\sigma_{\mu\nu}(1-\gamma_5)/2\cdot(\lambda^a/2)$
and $f_{abc}$ and $d_{abc}$ are SU(3) symbols.
For anti-instanton replace $1-\gamma_5$ to $1+\gamma_5$.
Then the tunneling rate at finite density is given by
the average of \eq{efflag} over the ground state of nuclear
matter:
\ben
n(\rho,\rho_B)=\lgl \Delta{\cal L}\rgl_{\rho_B}. \label{tunrate-rhoB1}
\een
The expectation values of multi quark operators in
\eq{tunrate-rhoB1}
are evaluated by applying factorization hypothesis.
In this approximation, all the terms in \eq{efflag}
containing $\gamma_5$, $\sigma_{\mu\nu}$, $\lambda^{a}/2$
drop off [\ref{Shifman}].
As a result, \eq{tunrate-rhoB1} 
is reduced to the same form as \eq{tunrate-nv} but with the current quark 
mass replaced by the \lq\lq effective mass'':
\ben
n(\rho,\rho_B)=n_{0}(\rho)
(\rho\mu_0)^{n_f \gamma_0 \frac{\alpha_{\msbar}(\mu_0)}{4\pi}}
\prod_{q=u,d,s}m_{q}^{*}(\rho,\rho_B)\rho,\label{tunrate-rhoB2}
\een
where the \lq\lq effective mass'' is defined by
\ben
m_q^{*}(\rho,\rho_B)=m_q-\frac{2}{3}\pi^2\rho^2
\langle{\bar q}q\rangle_{\rho_B}.
\een
In order to know the derivative of \eq{tunrate-rhoB2}
with respect to $\rho_B$ in \eq{Pi-inst},
we must know the $\rho_B$ dependence of $\langle{\bar q}q\rangle_{\rho_B}$
in the effective quark mass.
$\langle{\bar q}q\rangle_{\rho_B}$ is expanded in powers of $\rho_B$ as 
$\langle{\bar q}q\rangle_{\rho_B}=\langle{\bar q}q\rangle_{0}
+\frac{\sgmtrm}{m_u+m_d}\rho_B+\cdots$ for $q=u,d$ and
$\langle{\bar s}s\rangle_{\rho_B}=\langle{\bar s}s\rangle_{0}
+y\frac{\sgmtrm}{m_u+m_d}\rho_B+\cdots$ [\ref{HL}], 
where $y=2\N{s}/(\N{u}+\N{d})$
is the strangeness content of the nucleon.
Then we obtain the derivative of \eq{tunrate-rhoB2}
with respect to $\rho_B$ as
\begin{eqnarray}
\left[\frac{\partial}{\partial\rho_B}n(\rho,\rho_B)\right]_{\rho_B=0} 
&=&n_{0}(\rho)
(\rho\mu_0)^{n_f \gamma_0 \frac{\alpha_{\msbar}(\mu_0)}{4\pi}}
\frac{-2\pi^2\rho^5}{3}\frac{\sgmtrm}{m_u+m_d}\cr
&&\times
\left[m_u^*(\rho,0) m_s^*(\rho,0) +m_d^*(\rho,0) m_s^*(\rho,0) +ym_u^*(\rho,0) m_d^*(\rho,0)\right]. \cr
&&\label{tunrate-N}
\end{eqnarray}

Now we have all the ingredients for deriving the QCD sum rule.
Collecting all the terms in \eq{contract-Pi}, namely,
the one-loop term \eq{OPE0} and the two-loop
term \eq{Pi-inst} with \eq{tunrate-N},
and substituting \eq{contract-Pi} into \eq{BSR0}, 
we obtain the QCD sum rule for 
$\gA{0}$ as follows:
\bey
|\gA{0}|^2&=&-\frac{M}{3}\left\{
\frac{\sgmtrm}{s}\left[\frac{28}{3}\left(1-{\m{s}\over\m{u}+\m{d}}\right)\right]
+\frac{\Sigma_{KN}}{s}\left[\frac{56}{3}{\m{s}\over\m{s}+\m{u}}\right]
\right. \nnbr \\
&&
+\frac{M}{s}\left[\frac{2}{3}-7\left(\mmnt{2}{u}
+\mmnt{2}{d}+\mmnt{2}{s}\right)\right]
\nnbr \\
&&
-\frac{4\pi\alpha_s\qqv}{s^2}
 \left[\frac{352}{27}\frac{\sgmtrm}{\m{u}+\m{d}}\right]
-\frac{4\pi\alpha_s\ssv}{s^2}
\left[\frac{176}{27}\left(\frac{2\Sigma_{KN}}{\m{s}+\m{u}}-\frac{\sgmtrm}{\m{u}+
\m{d}}\right)\right]\cr
&&\left.
+\frac{15M^3}{s^2}\left[\mmnt{4}{u}+\mmnt{4}{d}+\mmnt{4}{s}\right]
+I(s)
\right\},  \label{QSR0I} 
\eey
where $\qqv\equiv \uuv=\ddv$. 
In \eq{QSR0I} we assume $m_u=m_d$.
$I(s)$ is the instanton contribution which is given by
\ben
I(s)=9\int^{\rho_c}_0 d\rho
\left[\frac{\partial}{\partial\rho_B}n(\rho,\rho_B)\right]_{\rho_B=0} 
\rho^4
\int_{0}^{\infty} dt(s-\rho^2 {\rm cosh}^{2}t\cdot s^2)
{\rm exp}(-\rho^2 {\rm cosh}^{2}t \cdot s).
\een
Here the integration over $\rho$ has been cut off at the critical size
$\rho_c$.
We note that the value of $\rho_c$ estimated by Shuryak
is close to the upper boundary for the validity of the expansion \eq{efflag} \Ref{Shifman}.

We show in Fig.\ref{fig1} the square of the Borel mass, $s$, dependence of $|\gA{0}|$ in
\eq{QSR0I}.
In plotting the curve in Fig.\ref{fig1}, we used the following values
of the constants in the right hand side in \eq{QSR0I}.
The $\pi$-$N$ sigma-term is taken from Ref.[\ref{GLM}], which are $\sgmtrm 
=45\,{\rm MeV}$. 
The quark masses are taken to be $\m{u}=\m{d}=7\,{\rm MeV}$, 
$\m{s}=110\,{\rm MeV}$ \Ref{yazaki}.
Using the above values and the ratio 
$y=2\N{s}/(\N{u}+\N{d})=0.2$ given in Ref.[\ref{GLM}], we can calculate the $K$-$N$ sigma term averaged over the iso-spin states 
and the result is $\Sigma_{KN}=226\,{\rm MeV}$. 
We calculated the moments of parton
distributions adopting the LO scheme in Ref.\Ref{Gluck}: 
$\mmntQSR{2}{u}+\mmntQSR{2}{d}= 1.1$, 
$\mmntQSR{4}{u}+\mmntQSR{4}{d}= 0.13$,
$\mmntQSR{2}{s}= 0.03$, $\mmntQSR{4}{s}= 0.002$.
The vacuum condensates are taken from Ref.\Ref{yazaki}, 
which are $\qqv=(-225\,{\rm 
MeV})^3$ and $\ssv=0.8\qqv$.
The normalization point $\mu_0$ in \eq{tunrate-rhoB2}
was taken to be $1\,{\rm GeV}$, which is the relevant scale for 
QCD sum rules.
For the critical size of an (anti-)instanton we used the value
$\rho_c=0.3\,{\rm fm}$.
The upper curve in Fig.\ref{fig2} corresponds to $\gA{0}$ in \eq{QSR0I} but without
the instanton contribution $I(s)$.
We see that the curve is well stabilized.
In the stabilized region $\gA{0}\simeq 0.8$.
This is consistent with the well known fact that $\gA{0}$ 
is about $30\%$ suppressed due to relativistic effect
compared with the naive quark model's expectation,
but is much larger than the experimental value $\gA{0}=(0.28-0.41)$.
The lower three curves correspond to $\gA{0}$
including the instanton contribution $I(s)$
with different choices of $\rho_c$.
We see the Borel curve is extremely sensitive to $\rho_c$ 
and not stabilized.
Therefore we cannot predict the value of $\gA{0}$.
However, we can say
that apparently instantons tend to lower $\gA{0}$ compared with that 
without the instanton contribution.

In summary, we have derived a QCD sum rule for 
$\gA{0}$ from a two-point correlation function
of flavour singlet axial-vector currents in one-nucleon state.
In deriving the sum rule, we evaluated the correlation function by an OPE
up to dimension 6.
We have also took into account an additional contribution arising
from an (anti-)instanton and evaluated it under DIGA.
When we do not include the instanton contribution,
$\gA{0}$ is not so suppressed as the experimental value and is about 
$0.8$.
Including the instanton contribution $\gA{0}$
tends to be suppressed compared with the result when we do not
include the instanton contribution.
Recently, Sch\"afer and Zetocha \Ref{Schafer} have computed the axial coupling constants
of the nucleon using numerical simulations of the instanton liquid.
They found the isovector axial coupling constant is $\gA{3}=1.28$,
in good agreement with experiment,
while flavour singlet coupling is $\gA{0}=0.77$.
$\gA{0}$ comes from a connected part and OZI violating disconnected part
of the three point correlation function.
Taking into account only the connected part they found $\gA{0}=0.79$,
while the disconnected part is very small, $\gA{0}(dis)=-(0.02\pm0.02)$.
It would be interesting if we can clarify the relation between the present result obtained
by adding a single instanton contribution to the OPE-based QCD sum rules
and that by the instanton liquid model.

The author would like to thank 
Professor Y. Kondo and Professor S. Saito
for valuable discussions.
I especially thank Professor M. Oka
for helpful discussions and careful reading of the manuscript. 

\newpage
\baselineskip 24pt
\begin{center}
{\bf References}
\end{center}
\def\labelenumi{[\theenumi]}
\def\Ref#1{[\ref{#1}]}
\def\Refs#1#2{[\ref{#1},\ref{#2}]}
\def\anp#1#2#3{{Adv. Nucl. Phys.\,}{\bf{#1}}\,(#3),#2}
\def\npb#1#2#3{{Nucl. Phys.\,}{\bf B{#1}}\,(#3),#2}
\def\npa#1#2#3{{Nucl. Phys.\,}{\bf A{#1}}\,(#3),#2}
\def\np#1#2#3{{Nucl. Phys.\,}{\bf{#1}}\,(#3),#2}
\def\plb#1#2#3{{Phys. Lett.\,}{\bf B{#1}}\,(#3),#2}
\def\prl#1#2#3{{Phys. Rev. Lett.\,}{\bf{#1}}\,(#3),#2}
\def\prd#1#2#3{{Phys. Rev.\,}{\bf D{#1}}\,(#3),#2}
\def\prc#1#2#3{{Phys. Rev.\,}{\bf C{#1}}\,(#3),#2}
\def\pr#1#2#3{{Phys. Rev.\,}{\bf{#1}}\,(#3),#2}
\def\ap#1#2#3{{Ann. Phys.\,}{\bf{#1}}\,(#3),#2}
\def\prep#1#2#3{{Phys. Reports\,}{\bf{#1}}\,(#3),#2}
\def\rmp#1#2#3{{Rev. Mod. Phys.\,}{\bf{#1}}\,(#3),#2}
\def\cmp#1#2#3{{Comm. Math. Phys.\,}{\bf{#1}}\,(#3),#2}
\def\ptp#1#2#3{{Prog. Theor. Phys.\,}{\bf{#1}}\,(#3),#2}
\def\ib#1#2#3{{\it ibid.\,}{\bf{#1}}\,(#3),#2}
\def\zsc#1#2#3{{Z. Phys. \,}{\bf C{#1}}\,(#3),#2}
\def\zsa#1#2#3{{Z. Phys. \,}{\bf A{#1}}\,(#3),#2}
\def\intj#1#2#3{{Int. J. Mod. Phys.\,}{\bf A{#1}}\,(#3),#2}
\def\sjnp#1#2#3{{Sov. J. Nucl. Phys.\,}{\bf #1}\,(#3),#2}
\def\pan#1#2#3{{Phys. Atom. Nucl.\,}{\bf #1}\,(#3),#2}
\def\app#1#2#3{{Acta. Phys. Pol.\,}{\bf #1}\,(#3),#2}
\def\etal{{\it et al.}}
\begin{enumerate}
\divide\baselineskip by 4
\multiply\baselineskip by 3
\item \label{Anselmino}
M. Anselmino, A. Efremov and E. Leader, \prep{261}{1}{1995}.
\item \label{Ji}
B.W. Filippone and X.D. Ji, \anp{26}{1}{2001}.
\item \label{SMC1}
SMC, B. Adeva et al., \plb{302}{533}{1993}; 
D. Adams et al.,
\plb{329}{399}{1994}; 
D. Adams et al., \plb{339}{332(E)}{1994};
D. Adams et al., \plb{357}{248}{1995}; 
B. Adeva et al.,
\plb{369}{93}{1996}.
\item \label{SLAC1}
SLAC-E142 Collab., P. L. Antony et al., \prl{71}{959}{1993}.
\item \label{SLAC2}
SLAC-E143 Collab., K. Abe et al., \prl{75}{25}{1995};
\prl{74}{346}{1995}; \prl{76}{587}{1996}.
\item \label{SMC2}
SMC, D. Adams et al., \prd{56}{5330}{1997}.
\item \label{SLAC3}
SLAC-E154 Collab., K. Abe et al., \plb{405}{180}{1997}.
\item \label{Ioffe}
B. L. Ioffe and A. G. Oganesian, \prd{57}{R6590}{1998}.
\item \label{Belitsky}
A. V. Belitsky and O. V. Teryaev, \pan{60}{455}{1997}.
\item \label{Balitsky}
Ya. Ya. Balitsky, A. V. Kolesnichenko and A. V. Yung, \sjnp{41}{178}{1985}.
\item \label{TN}
T. Nishikawa, S. Saito and Y. Kondo, \prl{84}{2326}{2000}.
\item \label{yazaki}
L. J. Reinders, H. R. Rubinstein, and S. Yazaki,
\prep{127}{1}{1985}.
\item \label{SVZ1} M. A. Shifman, A. I. Vainshtein and~V. I. Zakharov, 
\npb{147}{385}{1979};448.
\item \label{GLM} J. Gasser, H. Leutwyler and M. E. Sainio,
\plb{253}{252}{1991}.
\item \label{HL}
T. Hatsuda and S. H. Lee, \prc{34}{R46}{1992}.
\item \label{Gluck} M. Gl\"uck, E. Reya and A. Vogt,
\zsc{53}{127}{1992}.
\item \label{Shuryak}
T. Schaefer and E. V. Shuryak, \rmp{70}{323}{1998}.
\item \label{Shuryak-orgnl}
E. V. Shuryak, \npb{203}{93}{1982}; {\bf B203}\,(1982),116.
\item \label{tHooft}
G. 'tHooft, \prd{14}{3432}{1976}.
\item \label{Schrempp}
A.~Ringwald and F.~Schrempp, \plb{459}{249}{1999}.
\item \label{Andrei}
N. Andrei and D. J. Gross, \prd{18}{468}{1978}.
\item \label{Brown}
L. S. Brown, R. D. Carlitz, D. B. Creamer and C. Lee,
      \prd{17}{1583}{1978}.
\item \label{Din}
A. M. Din, F. Finjord and W. J. Zakrzewski, \npb{153}{46}{1979}.
\item \label{Shifman}
M. A. Shifman, A. I. Vainshtein and~V. I. Zakharov, 
\npb{163}{46}{1980}.
\item \label{Schafer}
T. Sch\"afer and V. Zetocha, \prd{69}{094028}{2004}.
\end{enumerate}
\clearpage
\begin{figure}[h]
\begin{center}
\includegraphics[width=13cm,keepaspectratio]{fig1.eps} 
\end{center}
\caption{(a)
Zero-modes propagating through an instanton.
The solid lines correspond to zero-modes 
with the chirality left-handed (L) and 
right-handed (R).
An instanton is shown as an open circle with \lq\lq I''.
A zero-mode changes its chirality when passing through an instanton.
(b) The diagram corresponding to the two-loop term, \eq{2loop}.
Quarks created by an axial current have same chirality.
So quarks do not propagate in an zero-mode state,
which changes its chirality in passing through an instanton.}
\label{fig1}
\end{figure}
\clearpage
\pagestyle{empty}
\begin{figure}[h]
\begin{center}
\includegraphics[width=15cm,keepaspectratio]{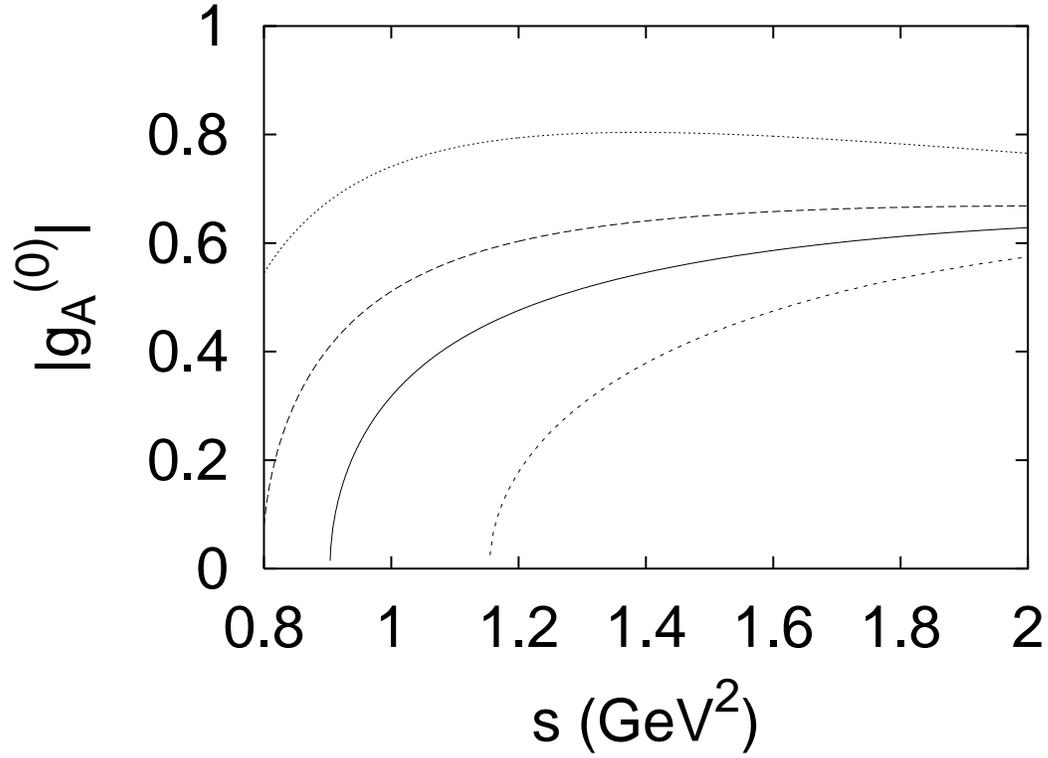} 
\end{center}
\caption{The square of the Borel mass, $s$, dependence of $|\gA{0}|$ in \eq{QSR0I}
with and without the instanton contribution.
The upper curve (long-dashed line) corresponds to that without the instanton contribution.
The lower three curves show those with the instanton contribution
for different choices of $\rho_c$;
The solid line corresponds to $\rho_c=0.3\,{\rm fm}$,
the long-dashed line to $\rho_c=0.29\,{\rm fm}$ and the short-dashed line to $\rho_c=0.31\,{\rm fm}$. }
\label{fig2}
\end{figure}
\end{document}